\documentclass[reprint,
superscriptaddress,
amsmath,
amssymb,
aps,
two columns,
fleqn,
floatfix
]{revtex4-1} %REVTEX
\usepackage{verbatim}
\usepackage{graphicx}
\usepackage{dcolumn}
\usepackage{bm}
\usepackage{mathtools}

\usepackage{graphicx}
\usepackage{hyperref}
\usepackage{multirow}
\usepackage{hhline}                         
\usepackage[usenames, dvipsnames]{color}
\usepackage{amssymb, amsfonts}
\usepackage{booktabs}
\usepackage[ruled]{algorithm2e}
\usepackage{algcompatible}
\usepackage{cancel}

\newcolumntype{P}[1]{>{\centering\arraybackslash}p{#1}}

\usepackage{soul}

\usepackage{lipsum}

\graphicspath{{figures_main}}
\usepackage{color}

\begin{document}

\title{The bliss of dimensionality: how an unsupervised criterion identifies\\ optimal low-resolution representations of high-dimensional datasets}

\author{Margherita Mele}
\affiliation{Physics Department, University of Trento, via Sommarive, 14 I-38123 Trento, Italy}
\affiliation{INFN-TIFPA, Trento Institute for Fundamental Physics and Applications, I-38123 Trento, Italy}

\author{Daniel Campos Moreno}
\affiliation{Grup de F\'isica Estadística, Departament de F\'isica, Facultat de Ci\`encies, Universitat Aut\`onoma de Barcelona, 08193 Bellaterra, Barcelona, Spain}

\author{Raffaello Potestio}
\email{raffaello.potestio@unitn.it}
\affiliation{Physics Department, University of Trento, via Sommarive, 14 I-38123 Trento, Italy}
\affiliation{INFN-TIFPA, Trento Institute for Fundamental Physics and Applications, I-38123 Trento, Italy}

\date{\today}

\begin{abstract}
Selecting the optimal resolution for discretizing high-dimensional data is a central problem in physics and data analysis, particularly in unsupervised settings where the underlying distribution is unknown. The resolution--relevance (Res--Rel) framework addresses this issue through an information-theoretic trade-off between descriptive detail and statistical reliability. Here we provide a systematic validation of this approach by comparing its characteristic optima --- maximum relevance and the $-1$ slope point --- with the discretization that minimizes the Kullback--Leibler divergence from a known or physically motivated ground truth distribution. Across unstructured and structured synthetic datasets, Gaussian clones of MNIST, and molecular dynamics simulations of the alanine dipeptide, we find that as the dimensionality or informative content increases the KL-optimal discretization consistently lies within the Res--Rel optimality region. Furthermore, in high-dimensional regimes the $-1$ slope criterion closely matches the KL divergence minimum. These results establish the quantitative consistency of unsupervised information-theoretic selection with distribution-based optimality.
\end{abstract}

\maketitle

\section{Introduction}

Discrete representations of continuous data cover a wide range of situations, from building simple one-dimensional histograms to complex clustering techniques for high-dimensional datasets. Such methods are ubiquitous across physics \cite{Kadanoff1966}, data science \cite{Jain1999}, and machine learning \cite{Vapnik1998}, and provide the bedrock for tasks such as coarse-graining \cite{Wilson1974}, density estimation \cite{Silverman2018}, and model inference \cite{Akaike1974}. 
Here, as in many other related contexts, choosing the appropriate level of detail of the representation remains a fundamental open problem \cite{Rissanen1978,VapnikChervonenkis2015}: indeed, overly coarse descriptions discard important structure, whereas excessively fine ones introduce sampling noise and statistical unreliability \cite{BickelLevina2004,Wainwright2019}.
This trade-off is particularly acute in finite-sample regimes and high-dimensional systems, where empirical frequencies may poorly approximate the underlying generative process of the data \cite{Bellman1966,DevroyeGyorgfiLugosi2013,Belkin2019}.
Standard criteria developed to select an adequate level of detail of the data representation often rely on supervised information about the target distribution (e.g., likelihood optimisation or divergence minimisation) \cite{Gelman2014,Schwarz1978,BurnhamAnderson2004}, but such an approach is naturally excluded in unsupervised settings.
Consequently, intrinsically data-driven criteria are needed to identify informative representations without explicit knowledge of the true probability distribution \cite{TishbyZaslavsky2015,Jaiswal2021,LeCun2015}.
The resolution--relevance (Res--Rel) framework \cite{Marsili2013,HaimoviciMarsili2015} addresses this problem by introducing an information-theoretic criterion that balances representational granularity (resolution) against statistical significance (relevance). In this approach, clusters of instances of a dataset are grouped together according to a given criterion, thereby inducing a spectrum of reduced representations of the data that are evaluated in terms of their resolution and relevance; by analysing how empirical frequency statistics changes with the number of discrete states, the framework identifies optimal resolution regimes directly from the data, without external supervision \cite{MarsiliRoudi2022}.
The simplicity and the unsupervised character of the framework make it extremely attractive and potentially powerful, and its increasingly numerous applications provide concrete support to its effectiveness \cite{Grigolon2016,Battistin2017, Holtzman2022, Mele2022, Morand2024, Aldrigo2025, MeleFiorentiniTarenzi2026}; furthermore, an important body of work has been done to lay solid mathematical and statistical foundations to the approach \cite{Song2018,Cubero2018,Cubero2019,Duranthon2021}. However, a simple, bottom-up validation of its working and its range of validity is still missing. In this letter, we address this problem by analysing synthetic data points generated from distributions known \emph{a priori}, as well as realistic datasets such as MNIST and molecular dynamics trajectories of the alanine dipeptide, systematically evaluating the capability of the resolution--relevance approach to provide physically informative low-resolution representations of high-dimensional data and setting boundaries for its applicability.

\section{Methods}

Consider a $N$-dimensional dataset of size $P$ partitioned into $n$ discrete states, or clusters, by a mapping that assigns each data point to a state. Let $k_i$ denote the number of data points assigned to state $i$, and $m_k$ the number of states with occupancy exactly equal to $k$. Resolution and relevance are defined, respectively, as:
\begin{equation} 
H_{\mathrm{res}} = - \sum_{i=1}^{n} \frac{k_i}{P}  \log \frac{k_i}{P}, \; H_{\mathrm{rel}} = - \sum_{k} \frac{k m_k}{P}  \log \frac{k m_k}{P}.
\end{equation}

Resolution, $H_{\mathrm{res}}$, is the Shannon entropy of the empirical distribution of frequencies, and quantifies the level of detail of the representation. Relevance, $H_{\mathrm{rel}}$, instead captures the heterogeneity of empirical frequencies through the occupancy distribution ${m_k}$, and so reflects the amount of statistically significant information entailed in the coarse representation of the data \cite{MarsiliRoudi2022,Grigolon2016}. By varying the number of states one can generate the \emph{resolution--relevance} curve, which shows and quantifies the trade-off between descriptive detail and statistical reliability \cite{Song2018}. Resolution increases with the number of states, while relevance displays a non-monotonic behaviour: it first increases as more and more informative structure is resolved, then decreases when further refinement produces poorly sampled states dominated by noise. An optimal representation thus lies in the intermediate regime, between the maximum relevance and the point where the curve attains slope $-1$, which marks the information-theoretical (IT) optimum \cite{Duranthon2021,HaimoviciMarsili2015}. We consider this interval comprised between the relevance maximum and the IT optimum to represent the \emph{optimality region}: beyond this point, information gains from increasing resolution are lower than the corresponding information losses by reducing statistical significance. 

Each representation induces an empirical probability distribution: given a partition $\{C_j\}_{j=1}^n$ with occupancies $k_j = |C_j|$, the empirical probability assigned to each data point is $\hat{p}(x_i) = {k_j}/{P}$, for $x_i \in C_j$. If the true generative distribution $p(x)$ were known, a representation at the optimal level of resolution would minimise the discrepancy between $p(x)$ and the empirical distribution $\hat{p}(x)$. A natural measure for this is the Kullback--Leibler (KL) divergence between both distributions,
\begin{equation}
D_{\mathrm{KL}}(p \| \hat{p}) = \sum_i p(x_i)\, \log \frac{p(x_i)}{\hat{p}(x_i)} .
\end{equation}

When evaluating the KL divergence, the reference probability density is evaluated at the data points and normalised over the dataset; the empirical representation is treated consistently, so that both induced discrete distributions are properly normalised, such that $\sum_i p(x_i) = \sum_i \hat{p}(x_i) = 1$. In typical unsupervised settings, however, the true distribution $p(x)$ is unknown and cannot be used to guide the choice of representation. The problem therefore reduces to identifying informative discretizations directly from empirical frequency statistics, and the resolution--relevance framework provides a fully data-driven criterion to select such representations. To assess how the representations selected by this criterion relate to optimal ones reproducing the empirical distribution, we analyse datasets for which such empirical is available--either known by construction (synthetic data) or physically/statistically motivated (real systems). This controlled setting enables a direct comparison between the Res--Rel optima and the discretisation that minimises the KL divergence from the reference. The datasets analysed and the associated analysis procedure are summarised below, with full technical details reported in the Supplemental Material (SM)~\cite{supplemental}.

\begin{figure}[t]
    \centering
    \includegraphics[width=1\linewidth]{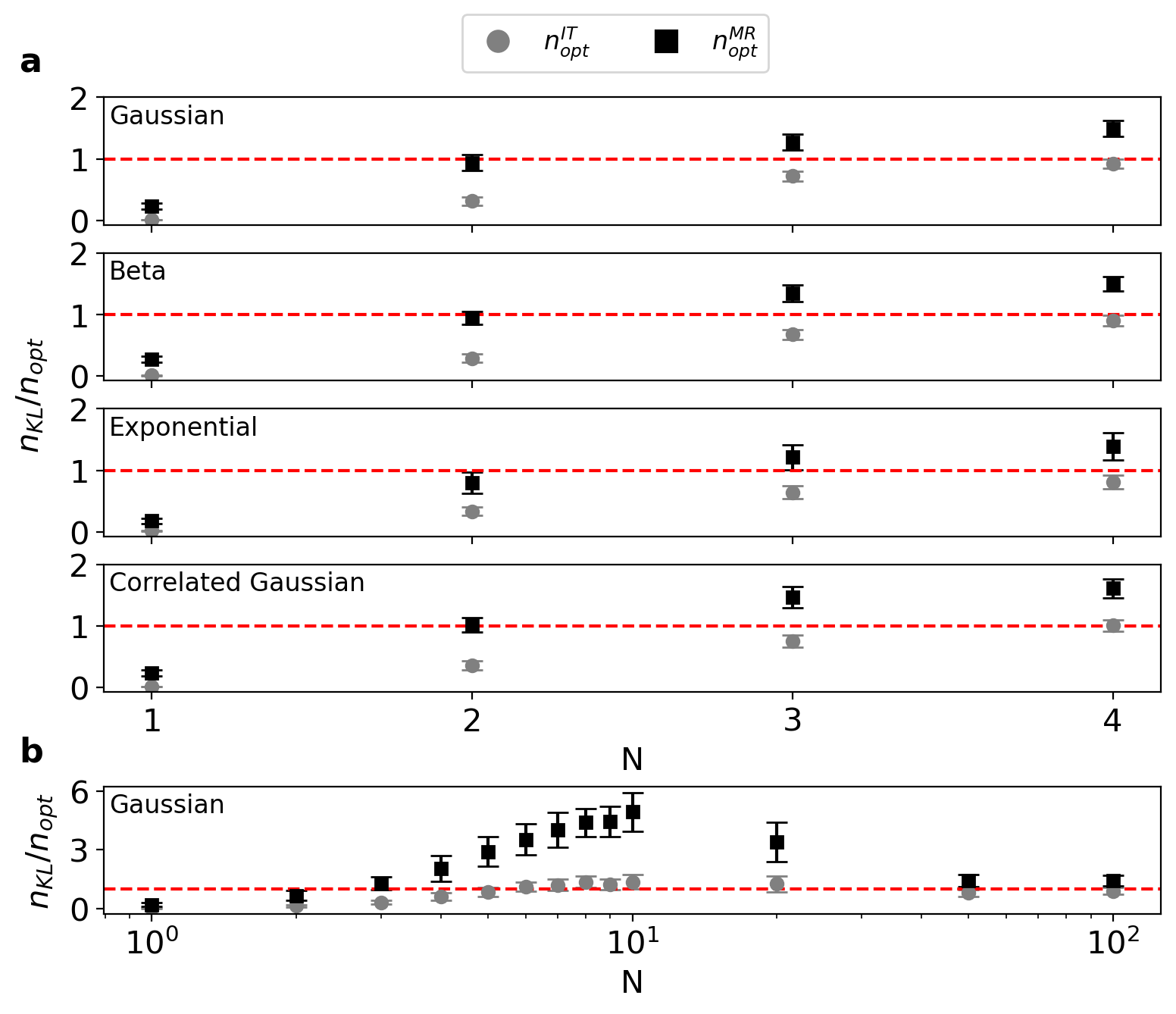}
    \caption{\textbf{Comparison between resolution--relevance and Kullback--Leibler optimal representations for unstructured synthetic data.} Ratio $n_{\mathrm{KL}}/n_{\mathrm{opt}}$ as a function of the data dimensionality $N$ for unstructured synthetic datasets. In each panel, black squares correspond to $n_{\mathrm{opt}}^{\mathrm{MR}}$ (maximum relevance), while grey circles correspond to $n_{\mathrm{opt}}^{\mathrm{IT}}$ ($-1$ slope point). The red dashed line indicates equality between the two estimates. Error bars represent the standard deviation over $50$ independent realisations. Panel (a) shows low-dimensional datasets analysed using $N$-dimensional histograms, generated from Gaussian, Beta, Exponential, and correlated Gaussian distributions. Panel (b) shows the corresponding analysis for IID Gaussian data spanning from low to high dimensions, where representations are constructed using UPGMA clustering.}
    \label{fig:syntetic_unstructured}
\end{figure}

\begin{figure*}
    \centering
    \includegraphics[width=\linewidth]{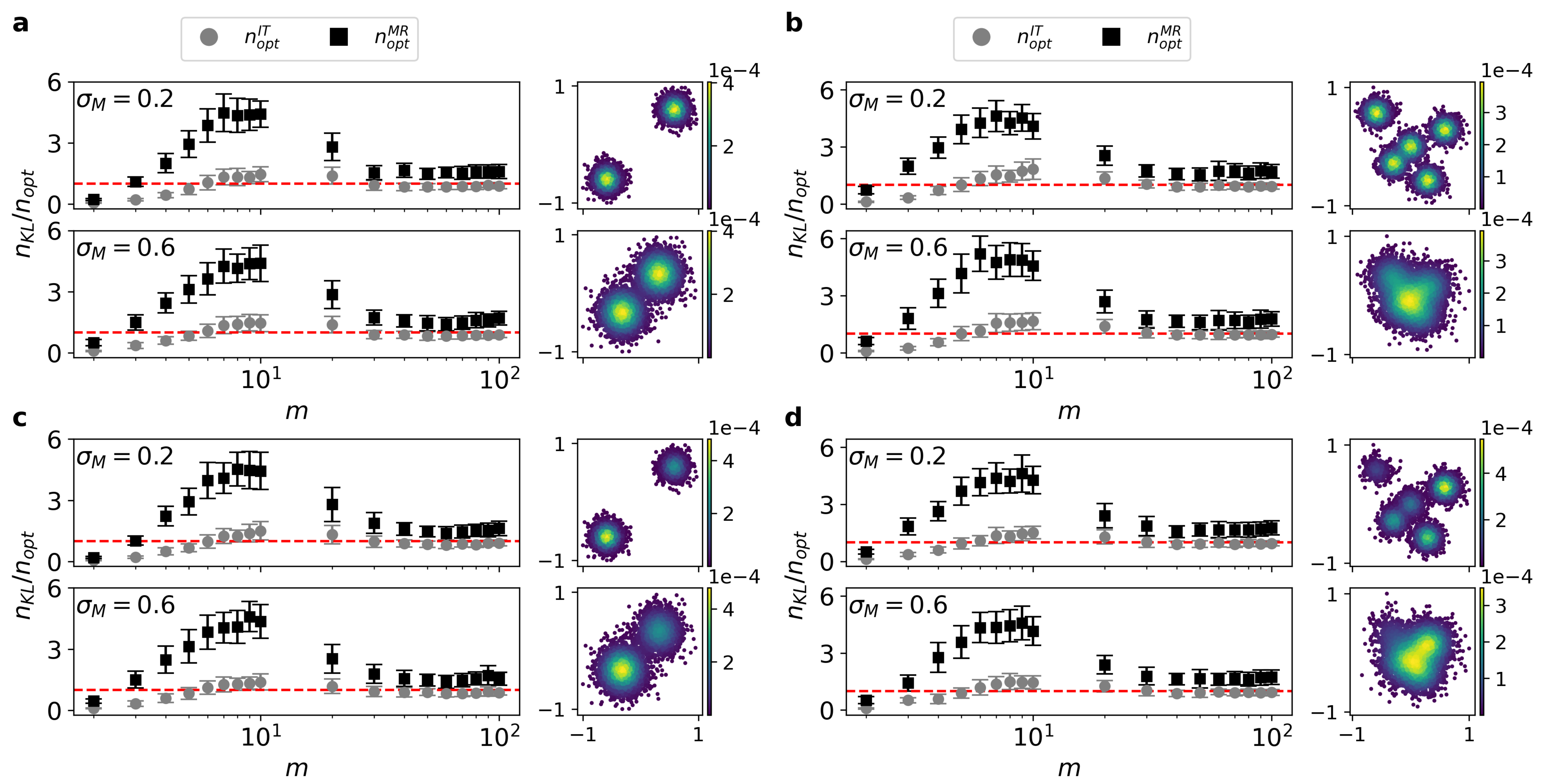}
    \caption{\textbf{Comparison between resolution--relevance and Kullback--Leibler optimal representations for structured synthetic data.}
    Ratio $n_{\mathrm{KL}}/n_{\mathrm{opt}}$ as a function of the number of informative dimensions $m$ for structured synthetic datasets with latent Gaussian mixture structure. In each panel, black squares correspond to $n_{\mathrm{opt}}^{\mathrm{MR}}$ (maximum relevance), while grey circles correspond to $n_{\mathrm{opt}}^{\mathrm{IT}}$ ($-1$ slope point). The red dashed line indicates equality between the two estimates. Error bars represent the standard deviation over $50$ independent realisations.
    Panels (a) and (b) show equal-weight mixtures for $K=2$ and $K=5$, while panels (c,d) show the corresponding unequal-weight cases with weights $[0.66,0.33]$ and $[0.34,0.27,0.19,0.13,0.07]$, respectively. Different rows correspond to increasing values of the mixture standard deviation $\sigma_{\mathrm{M}}$, as indicated.
    For each row, the right-hand subpanels show two-dimensional projections of representative datasets generated with $m=2$ for the corresponding parameter values. For each value of $K$, the random seed used to generate the data is fixed across different scenarios, ensuring that the relative positions of the mixture means are identical.}
    \label{fig:syntetic_structured}
\end{figure*}

\textit{Methodology ---}For each dataset realisation, representations are constructed by partitioning the data into $n$ discrete states. For a first test on low-dimensional ($N \leq 4$) datasets, we discretised the data space of all four distributions under examination using $N$-dimensional histograms. The direct discretization of spaces with $N > 4$ rapidly becomes numerically intractable (see below); hence, we focussed on the case of Gaussian distributions in spaces of dimension $1$ to $100$ partitioning them by agglomerative clustering with UPGMA linkage \cite{SokalMichener1958}. The number of states $n$ parametrises the representation and generates a unique resolution--relevance curve for each realisation. For every $n$ we compute $(H_{\mathrm{res}}, H_{\mathrm{rel}})$, and the empirical distribution induced by the partition. From each resolution--relevance curve we identify two characteristic values of $n$: the point of maximum relevance (MR), defining $n^{\mathrm{MR}}_{\mathrm{opt}}$, and the point where the curve attains slope $-1$, defining $n^{\mathrm{IT}}_{\mathrm{opt}}$. Independently, the KL divergence between the reference distribution and the empirical distribution is evaluated as a function of $n$, and the minimiser $n_{\mathrm{KL}}$ is identified. For synthetic and semi-real datasets, results are averaged over $50$ independent realisations; for the alanine dipeptide \cite{Smith1999} the analysis is performed over $10$ independent MD trajectories. Implementation details and additional robustness analyses are reported in the SM~\cite{supplemental}.

\begin{figure}
    \centering
    \includegraphics[width=1\linewidth]{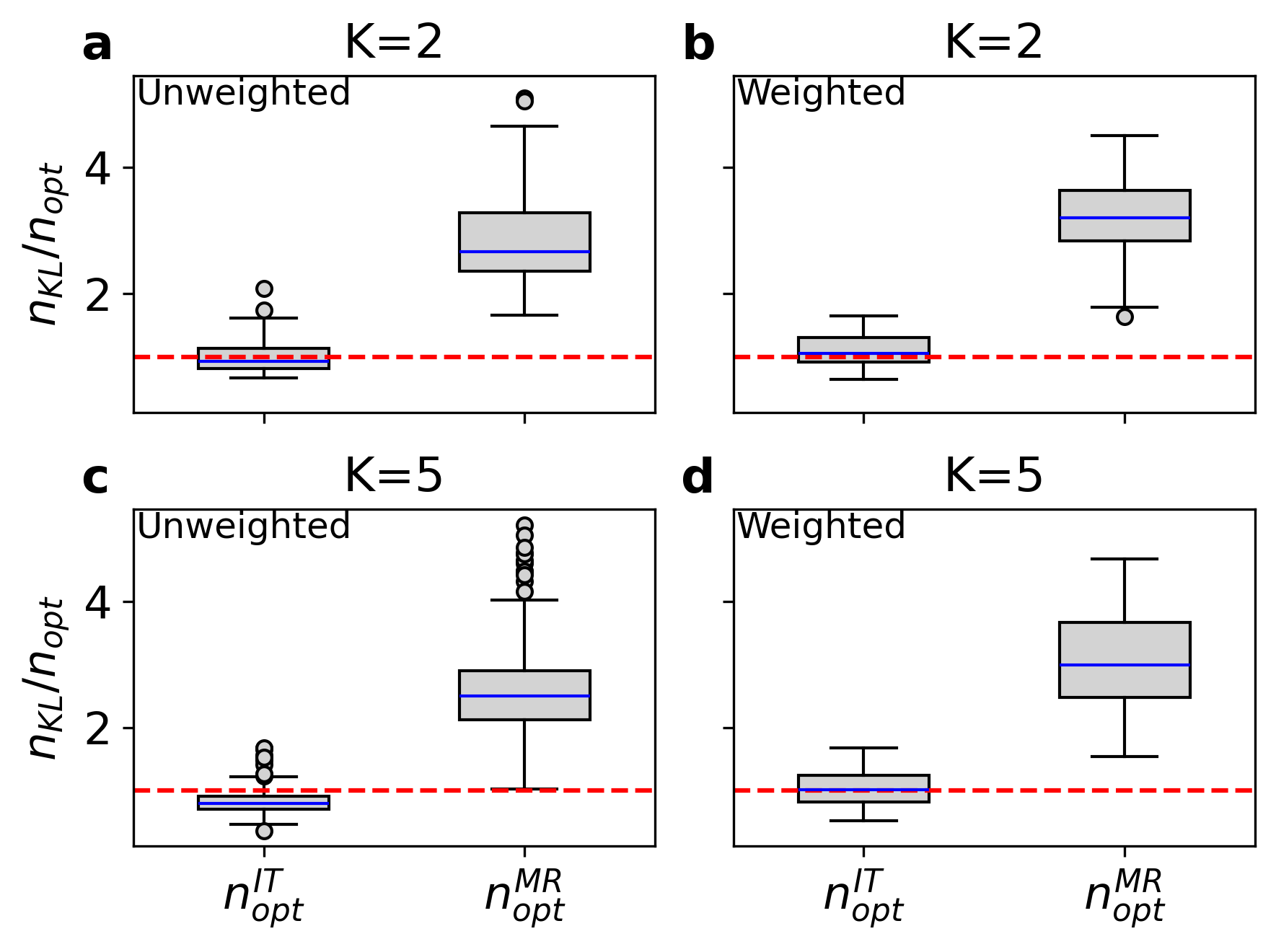}
    \caption{\textbf{Comparison between resolution--relevance and Kullback--Leibler optimal representations for Gaussian clones of MNIST.} Boxplots show the ratio \( n_{\mathrm{KL}} / n_{\mathrm{opt}} \) evaluated at two characteristic points of the resolution--relevance curve: the \(-1\)-slope point \( n_{\mathrm{opt}}^{\mathrm{IT}} \) (information-theoretic optimum) and the maximum-relevance point \( n_{\mathrm{opt}}^{\mathrm{MR}} \). The horizontal red dashed line marks the ideal value \( n_{\mathrm{KL}} / n_{\mathrm{opt}} = 1 \), corresponding to perfect agreement between the two criteria, while blue horizontal lines indicate the median of each distribution. Panels (a,b) correspond to mixtures with \( K = 2 \) components, with equal weights in (a) and unequal weights \( [0.66,\,0.33] \) in (b). Panels (c,d) show mixtures with \( K = 5 \) components, with equal weights in (c) and unequal weights \( [0.34,\,0.27,\,0.19,\,0.13,\,0.07] \) in (d).}
    \label{fig:MNIST}
\end{figure}

\section{Results}

\subsection{Unstructured Synthetic Data}

To establish a baseline, we first consider synthetic datasets without latent discrete structure, where all components are sampled independently from a prescribed continuous distribution; this setting allows us to isolate purely dimensional effects. Datasets are generated from Gaussian, beta, exponential, and correlated Gaussian distributions (see SM~\cite{supplemental} for technical details). For low-dimensional data ($N \leq 4$), representations are constructed \emph{via} $N$-dimensional histograms, which provide a direct and controlled discretisation of the data space and constitute the natural benchmark for estimating empirical probability densities. However, the number of bins grows exponentially with $N$, leading to severe sparsity and unstable frequency estimates at fixed sample size \cite{Bellman1966}. To probe higher-dimensional regimes, we therefore adopt agglomerative clustering (UPGMA), which defines discrete states directly from inter-point distances without relying on a regular grid. For each dimensionality $N$, we compare $n^{\mathrm{MR}}_{\mathrm{opt}}$ and $n^{\mathrm{IT}}_{\mathrm{opt}}$ with the KL-optimal value $n_{\mathrm{KL}}$. \autoref{fig:syntetic_unstructured}.a shows that in one dimension the Res--Rel framework overestimates the KL-optimal number of states, with both criteria yielding $n_{\mathrm{opt}} > n_{\mathrm{KL}}$. The discrepancy rapidly decreases with increasing $N$: already for $N=2$, $n^{\mathrm{MR}}_{\mathrm{opt}}$ is close to $n_{\mathrm{KL}}$, and for $N \geq 2$ the KL-optimal value systematically falls within the Res--Rel optimality region defined by $[n^{\mathrm{MR}}_{\mathrm{opt}}, n^{\mathrm{IT}}_{\mathrm{opt}}]$. This behaviour is robust across all tested distributions. The analysis is extended to higher dimensions using \emph{iid} Gaussian data (\autoref{fig:syntetic_unstructured}.b). Also in this case, the low-dimensional overestimation progressively vanishes as $N$ increases. For intermediate dimensions the Res--Rel optimality region broadens, while for $N > 10$ the two criteria converge, yielding close values of $n_{\mathrm{opt}}$. Overall, the discrepancy observed at low dimensionality progressively reduces as $N$ increases, and the KL-optimal value falls within the Res--Rel optimality region for $N \geq 2$.

\begin{figure}
    \centering
    \includegraphics[width=1\linewidth]{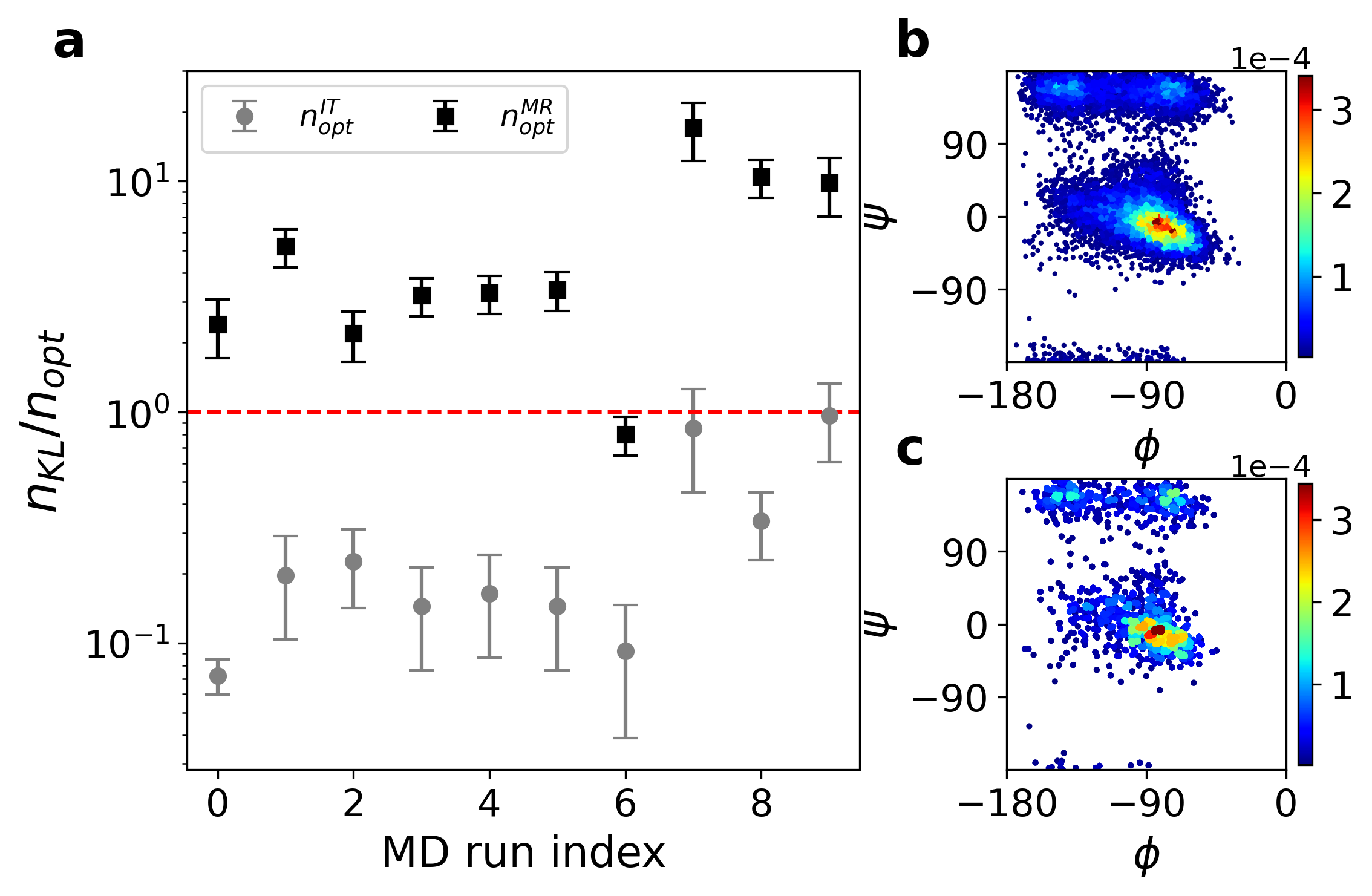}
    \caption{\textbf{Comparison between resolution--relevance and Kullback--Leibler optimal representations for Alanine dipeptide.} (a) Ratio \( n_{\mathrm{KL}} / n_{\mathrm{opt}} \) computed over ten independent molecular dynamics (MD) simulations. Grey circles denote the ratio evaluated at the maximum-relevance point \( n_{\mathrm{opt}}^{\mathrm{MR}} \), while black squares correspond to the information-theoretic optimum \( n_{\mathrm{opt}}^{\mathrm{IT}} \), defined as the \(-1\)-slope point of the resolution--relevance curve. The horizontal red dashed line marks the reference value \( n_{\mathrm{KL}} / n_{\mathrm{opt}} = 1 \), corresponding to perfect agreement between the Kullback--Leibler-optimal discretization and the resolution--relevance prediction. (b,c) Two-dimensional projections in the space of backbone dihedral angles \( (\phi, \psi) \) for a representative trajectory. In panel (b), colors represent a two-dimensional histogram of the raw MD frames, providing an empirical estimate of the reference probability distribution. In panel (c), only the centroids of the clusters obtained by RMSD-based clustering at \( n_{\mathrm{opt}}^{\mathrm{IT}} \) are shown. Colours are assigned from a two-dimensional histogram in \( (\phi,\psi) \) constructed over these centroids, each weighted by the multiplicity of its corresponding cluster.}
    \label{fig:alanin}
\end{figure}

\subsection{Structured Synthetic Data} 

Having characterised the dependence of the sheer dimensionality of the data, we now introduce latent discrete structure embedded in a high-dimensional space. The total number of dimensions is fixed to $N=100$, of which only $m \leq N$ components carry informative signal. In the informative subspace, samples are drawn from mixtures of $K$ Gaussian components with standard deviation $\sigma_M$, whereas the remaining dimensions consist of independent Gaussian noise with $\sigma << \sigma_M$ (see SM~\cite{supplemental} for full construction details). Both binary ($K=2$) and multi-component ($K=5$) mixtures, with equal and unequal weights, are analysed. \autoref{fig:syntetic_structured} reports the ratio between the KL-optimal number of discrete elements $n_{\mathrm{KL}}$ and the Res--Rel optima as a function of the number of informative dimensions $m$. For very low informative dimensionality ($m=2$), the Res--Rel criteria select values exceeding $n_{\mathrm{KL}}$, consistent with the behaviour observed in low-dimensional unstructured data. As $m$ increases, the agreement improves: the KL-optimal value systematically falls within the Res--Rel optimality region and is typically located close to the $-1$ slope point (grey circles). The width of the optimality region initially increases with $m$ and subsequently decreases for $m \geq 10$, as the informative signal dominates the noisy background. These trends are robust across all investigated mixture configurations and signal widths (additional parameter combinations are reported in SM~\cite{supplemental}).

\subsection{Semi-real Data: Gaussian Clones of MNIST}

To bridge fully synthetic models and real-world data, we analyse Gaussian clones derived from the MNIST handwritten digit database \cite{Deng2012}. For each selected digit class, a class-conditional multivariate Gaussian distribution is estimated from the original data and used to generate synthetic samples (``Gaussian clones''). The reference distribution is thus a Gaussian mixture with $K$ components corresponding to distinct digit classes (see SM~\cite{supplemental} for estimation details). Digit 1 is excluded due to its anomalous Res--Rel behaviour (see SM~\cite{supplemental}). We analyse mixtures with $K=2$ and $K=5$, considering both equal and unequal class weights. For each configuration, all possible combinations of $K$ digit classes are examined. \autoref{fig:MNIST} reports the ratio $n_{\mathrm{KL}}/n_{\mathrm{opt}}$ evaluated at the two Res--Rel characteristic points. Across all scenarios, the $-1$ slope criterion yields values closely aligned with the KL-minimising discretisation, with distributions centred around $n_{\mathrm{KL}}/n^{\mathrm{IT}}_{\mathrm{opt}} \simeq 1$. In contrast, the maximum-relevance criterion systematically selects smaller numbers of discrete elements, resulting in $n_{\mathrm{KL}}/n^{\mathrm{MR}}_{\mathrm{opt}} > 1$, albeit with deviations remaining within a factor smaller than four. The same qualitative behaviour is observed for both equal and unequal weight mixtures and across different class combinations, and is consistent with the features of the unstructured and structured synthetic data discussed in the previous sections, particularly in those cases where the effective dimensionality of the data is in the $4-10$ range. Interestingly, this result is line with previous works \cite{hein2005intrinsic,facco2017estimating,pope2021intrinsic} that estimated the effective dimension of the MNIST dataset in the range $12-14$: our data thus suggest that the pairs or $5$-tuples of clone digits are embedded in a space of effective dimension $O(10)$.

\subsection{Real Data: Alanine Dipeptide}

We finally consider a dataset consisting of molecular dynamics (MD) simulations of the alanine dipeptide, a standard benchmark system whose equilibrium behaviour is well described in terms of the backbone dihedral angles $(\phi,\psi)$ \cite{Smith1999, Bolhuis2000}. As commonly done, the reference probability distribution is defined in this reduced space and estimated from each trajectory \emph{via} a two-dimensional histogram of the dihedral angles (see SM~\cite{supplemental} for simulation details). Discrete representations are instead constructed by clustering molecular configurations in the full configurational space using the atom-wise root mean square distance (RMSD) as a metric. This setup allows us to test whether the Res--Rel-selected discretisations recover the physically relevant probability landscape defined in dihedral angle space. \autoref{fig:alanin}.a reports, for ten independent trajectories, the ratio between the KL-optimal number of clusters $n_{\mathrm{KL}}$ and the Res--Rel optima. In all cases, $n_{\mathrm{KL}}$ lies within the optimality region. However, unlike the synthetic and semi-real datasets, no single Res--Rel criterion systematically coincides with the KL minimum across trajectories. \autoref{fig:alanin}.b-c compare, for a representative trajectory, the empirical dihedral angle density and the probability distribution reconstructed from clustering at $n^{\mathrm{IT}}_{\mathrm{opt}}$. The two representations display consistent large-scale conformational features. Despite the absence of an exact generative distribution and the finite-sampling nature of the reference estimate, the Res--Rel framework restricts the optimal discretisation to a narrow range of cluster numbers across independent trajectories, with a dispersion in line with what was observed in the previously discussed cases for a low number of informative features.

\section{Discussion}

In this work we carried out a systematic investigation of low-resolution representations of synthetic, semi-real, and real datasets, comparing the optimal clusterings obtained through the application of the resolution--relevance framework against those that minimise the KL divergence from the underlying generative model. In low-dimensional regimes, the Res--Rel criteria tend to overestimate the KL minimum; as the effective dimensionality or informative content of the data increases, however, the KL-optimal number of discrete elements falls within the Res--Rel optimality region. In high-dimensional settings, the $-1$ slope criterion provides the closest agreement with the KL divergence minimum: indeed, as already observed by Cubero and coworkers \cite{Cubero2019}, this point strikes an optimal balance between reducing resolution and maximising informativeness, and corresponds to the threshold between lossy and lossless compression; in contrast, the point of largest relevance is deep in the lossy compression regime, thereby leading to a potentially excessive coarsening. Crucially, also in a real molecular system, where the reference distribution is empirically estimated, the KL-based resolution level falls within the Res--Rel optimality region. However, even though the $-1$ slope point is closer to the reference than the relevance maximum, we observe an important dispersion among different simulation replicas; this shows that, in real case scenarios, it is advisable to consider the whole range of resolution in between the two optimal points $n_{\mathrm{opt}}^{\mathrm{IT}}$ and $n_{\mathrm{opt}}^{\mathrm{MR}}$. A central result of this work is that, for a given sample size, the larger the dimensionality of the embedding space of the data under examination, the easier it becomes for the Res--Rel analysis approach to identify the partition that most faithfully reproduces the key features of the underlying generative process. This behaviour is particularly relevant as the so-called \emph{curse of dimensionality} \cite{altman2018curse} here becomes a ``bliss'', reminiscent of the simplification ushered in by the thermodynamic limit in the context of statistical mechanics. In conclusion, the Res--Rel framework offers a principled, fully data-driven mechanism for selecting informative low-resolution representations of high-dimensional datasets without requiring explicit knowledge of the underlying generative distribution. Importantly, one of the most critical advantages of the approach, namely that it relies only on the occurrence statistics, requires that one interprets the results in the light of the clustering framework employed \cite{Cubero2019,Mele2022}; this, however, does not represent a limitation, but rather a inherent necessity for context in the application of an unsupervised analysis approach. Taken together, this investigation shows that the Res--Rel criterion identifies discretisations that are statistically robust and probabilistically meaningful, effectively linking unsupervised information-theoretic selection to supervised distribution-based optimality.

\section{Acknowledgements}
%The authors are indebted with \revise{XXX} for an insightful reading of the manuscript and useful comments.
RP acknowledges support from ICSC - Centro Nazionale di Ricerca in HPC, Big Data and Quantum Computing, funded by the European Union under NextGenerationEU. Views and opinions expressed are however those of the author(s) only and do not necessarily reflect those of the European Union or The European Research Executive Agency. Neither the European Union nor the granting authority can be held responsible for them.

This work was conducted in the spirit of the Slow Science Manifesto, advocating for collaborative and sustainable research (slow-science.com).

\section{Author contributions}

All authors conceived the study and proposed the method. MM produced the data and performed the analysis. RP and MM interpreted the results. All authors drafted the paper, reviewed the results, and approved the final version of the manuscript.

\section{Competing interests}

The authors declare no competing interests.

\section{Supporting information}
The Supplementary Information provides methodological details and additional analyses supporting the main text, including the definitions of relevance and resolution, and the description of the numerical procedure used to extract the optimal representation sizes, the construction of empirical and reference probabilities, the evaluation of $D_{\mathrm{KL}}(p\|\hat p)$, the protocols used to build discrete representations (histograms and UPGMA clustering), and complete dataset specifications. It also reports supplementary results and diagnostics for the MNIST digit ``1'', alanine-dipeptide visualisation, robustness and implementation checks, additional structured-synthetic parameter combinations, and KL-based analyses.

\section{Data availability}
The raw data associated with this work are freely available on Zenodo at \href{https://doi.org/10.5281/zenodo.18860257}{https://doi.org/10.5281/zenodo.18860257}.

\bibliographystyle{ieeetr}
\bibliography{biblio}

\end{document}